# Numerical modeling of in-plane thermal conductivity measurement methods based on a suspended membrane setup

Hanfu Wang[1,*], Yanjun Guo[1], Kaiwu Peng[1,3], Weiguo Chu[1,*], Guangming Chen[2,*]

[1] CAS Key Laboratory of Nanosystem and Hierarchical Fabrication，CAS Center of Excellence for Nanoscience，National Center for Nanoscience and Technology, Beijing 100190, P. R. China.

[2] College of Materials Science and Engineering, Shenzhen University, Shenzhen 518055, P. R. China.

[3] Sino-Danish College, Sino-Danish Center for Education and Research, University of Chinese Academy of Sciences, Beijing 100049, China

*Electronic mails: wanghf@nanoctr.cn (H. Wang) ; wgchu@nanoctr.cn (W. Chu); chengm@szu.edu.cn (G. Chen)

**Abstract:**

A numerical modeling study，based on 3D finite element method (FEM) simulation and 1D analytical solutions，has been carried out to evaluate the capabilities of two *ac* methods for measuring in-plane thermal conductivity of thin film deposited on the back of a suspended $SiN_x$ membrane setup. Two parallel metal strips are present on the top of the dielectric membrane. One strip (S1) serves as both heater and thermometer, while another one (S2) acts as thermometer only. For a modified phase shift (MPS) method, it is crucial to extract the in-plane thermal diffusivity from the phase shift of the temperature oscillation on S2. It was found that the frequency window for carrying out the data fitting became narrower as the in-plane thermal diffusivity of the composite membrane ($\alpha_{\parallel,C}$) increased, primarily due to the failure of the semi-infinite width assumption in the low frequency region. To ensure the validity of the method, the upper limit of $\alpha_{\parallel,C}$ should not exceed ~1.8 × 10$^{-5}$ m$^2$ s$^{-1}$ for the specific membrane dimension under consideration (1 × 1 mm$^2$). On the other hand, inspired by a modified Ångström method proposed by Zhu recently, we suggest a new data reduction methodology which takes advantage of the phase shift on both S1 and S2 as well as the amplitude on S1. Based on the simulation results, it is expected that the non-ideality associated with the three "observables" may be at least partially cancelled out. Therefore, the frequency window selection for carrying out the data fitting is not sensitive to the magnitude of $\alpha_{\parallel,C}$ and the upper limit of measurable $\alpha_{C,\parallel}$ can be further increased with respect to the MPS method. For typical specimen films whose in-plane thermal conductivity ranges from 0.84 W m$^{-1}$ K$^{-1}$ to 50 W m$^{-1}$ K$^{-1}$, the method proposed here yields a theoretical measurement uncertainty of less than 5%.

# 1. Introduction

Precise and reliable measurement of thermal conductivity of thin films is technically important for developing various kinds of electronic devices and components including micro-thermoelectric generators and coolers [1], thermal actuators [2], phase-change memories [3], and dielectric layers of the electronic circuits [4], and so on. It can also help researchers to deeply understand the conduction mechanisms in specific film structures [5]. Since many films are anisotropic in nature, it is often desirable to measure their thermal conductivity in both in-plane and cross-plane directions. There exists several well-established methods for measuring the cross-plane thermal conductivity ($\kappa_{\perp,f}$) such as the conventional 3ω method [6], the time-domain thermoreflectance method (TDTR) [7] and the frequency-domain thermoreflectance (FDTR) method [8]. Though these methods can be extended to extract the in-plane thermal conductivity ($\kappa_{\parallel,f}$) by coupling with two-dimensional (2D) models, it is believed that the non-linear propagation of the measurements error may bring about a large uncertainty to the final result [8, 9].

To date, most methods of measuring $\kappa_{\parallel,f}$ of submicron-thick films are based on one-dimensional (1D) or quasi-1D heat transport models. Among these methods, Völklein *et al.* have demonstrated a novel strategy that relies on a measurement chip integrated with two rectangular-shaped free-standing dielectric membranes of different dimensions on which the specimen films can be deposited. [10] A single metal line, which acts as both heater and thermometer, is created along the long axis of each membrane. The authors have taken into account the influence of the radiation heat loss from the membrane structure. By generating either a steady or an *ac* heating power on the metal strip and measuring corresponding temperature responses, the method is able to yield the in-plane thermal conductivities of the composite membrane and the bare dielectric membrane, from which $\kappa_{\parallel,f}$ of the test film can be derived.

Alternatively, the measurement chip may adopt a double-metal-strip (DMS) configuration in which one metal strip is periodically heated and the second one is used solely as a temperature sensor. The temperature rises on S1 and S2 can be acquired using the 3ω and 2ω voltage measurement techniques, respectively. [11] With this setup, Zhang and Grigoropoulos [12] have demonstrated a phase shift (PS) approach to measure the thermal diffusivity of 0.6 μm and 1.4 μm thick $SiN_x$ membranes from the temperature oscillation on the sensing strip. Bodenschatz *et al.* have characterized $\kappa_{\parallel,f}$ of 20–40 nm thick gold and platinum films by fitting the experimental data to a model that describes explicitly the heat diffusion in a two-layered geometry [11]. Ushirokita and Tada [13] recently utilized the DMS configuration to characterize $\kappa_{\parallel,f}$ of several 200–1000 nm thick conducting polymer films. In their work, the authors first measured the in-plane thermal diffusivity of the suspended membrane structure with the PS method, and used the obtained value together with the amplitude of the temperature oscillation on the heating strip to derive the in-plane thermal conductivity. This method is tentatively denoted as modified phase shift (MPS) method hereafter. The in-plane thermal conductivities of the polymer films (typically less than 1 W m$^{-1}$ K$^{-1}$) obtained in this way were comparable to literature values measured by other methods, which verify the reliability of the MPS method in measuring the low thermal conductivity thin films at room temperature.

In practical application, it is often necessary to carry out thermal conductivity characterization on the films with a thermal conductivity higher than 1 W m$^{-1}$ K$^{-1}$. It is still unclear to what extent that the MPS method can be applied to these samples. To further explore the application scope of the MPS method, we have performed a series of finite element method (FEM) simulations, and compared them with 1D analytical models. It is found that the method generally works fine if the in-plane thermal diffusivity of the composite membrane structure under test, $\alpha_{\parallel,C}$, is below ~ 1.8 ×10$^{-5}$ m$^2$ s$^{-1}$ for the specific membrane dimension of 1×1 mm$^2$, and the frequency window for carrying out the data fitting has been carefully selected.

Based on the FEM results and inspired by a modified Ångstrom method recently introduced by Zhu [14], here we propose a new data reduction methodology of extracting $\kappa_{\parallel,f}$. Unlike the MPS method, the membrane-based modified Ångström method (denoted as MMA method hereafter) has taken into account the phase shift of temperature oscillation on both metal strips, in addition to the amplitude on the heating strip. It was suggested by the simulation results that the non-ideality associated with the three "observables" may have been partially cancelled out during the data reduction process. The frequency window selection for carrying out the data fitting is therefore not sensitive to the magnitude of $\alpha_{\parallel,C}$. The upper limit of $\alpha_{C,\parallel}$ can be further increased with respect to the MPS method. For typical specimen films considered in this paper ($\kappa_{\parallel,f}$ ranges from 0.84 W m$^{-1}$ K$^{-1}$ to 50 W m$^{-1}$ K$^{-1}$), the newly proposed method yields a theoretical measurement uncertainty of less than 5%.

2. **Methods**

**2.1 1D analytical solutions**

The simulation model employed in this article is derived from the measurement chip used by Ushirokita and Tada [13] with some modifications (Fig. 1). In our model, a piece of 100 nm thick SiN$_x$ membrane is suspended over a silicon frame with a window size of 1 × 1 mm$^2$. On top of the membrane, two 5 um wide gold strips denoted as S1 and S2 are created in parallel. The strip S1 which coincides with one of central axis of the window serves as both a heater and thermometer, while S2 located 50 um away from S1 acts as a thermometer only. The lengths of S1 and S2 are 0.25 mm and 0.20 mm, respectively. The test film is deposited on the back of the SiN$_x$ membrane. Each metal strip is connected with two outer contact pads (I+ and I-) through current-injecting bridges and two inner contact pads (V+ and V-) by voltage-measurement bridges.

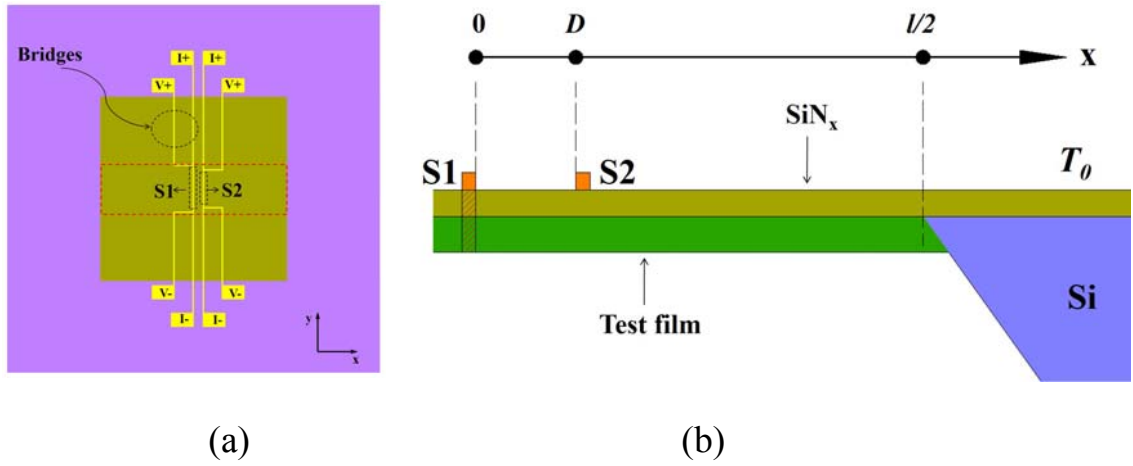

(a) (b)

Fig. 1 Top view (a) and side view (b) of the chip geometry used for modeling. The hatched area represents the membrane structure beneath S1, whose heat capacity is included in Eq. (8).

A sinusoidal current $I(t) = I_0 \sin(\omega t)$, where $I_0$ is the amplitude of the *ac* current, is injected into S1 from the outer contact pads to generate an oscillating heating power with a frequency of 2ω.:

$$P(t) = \frac{I_0^2 R_h}{2}\left(1-\cos(2\omega t)\right) = P_0 \left(1-\cos(2\omega t)\right) \qquad (1)$$

where $R_h$ is the resistance of S1 and $P_0$ is the amplitude of the heating power.

In the area surrounded by a red dashed square in Fig. 1a, the heat flow propagates approximately one-dimensionally along the +x and -x directions. The time dependent temperature oscillation $\widetilde{\Delta T}(x,t,2\omega)$ is thus given by the 1 D heat diffusion equation [14]:

$$\frac{1}{\alpha_{\|,j}}\frac{\partial \widetilde{\Delta T}(x,t,2\omega)}{\partial t} = \frac{\partial^2 \widetilde{\Delta T}(x,t,2\omega)}{\partial x^2} - \frac{8\varepsilon_j \gamma T_0^3}{d_j \kappa_{\|,j}}\widetilde{\Delta T}(x,t,2\omega) \qquad (2),$$

where $\kappa_{\|,j}$、$\alpha_{\|,j}$、$d_j$、$\varepsilon_j$ are the in-plane thermal conductivity, in-plane thermal diffusivity, thickness and emissivity of the membrane structure $j$ ($j=M,C$, $M$ represents the bare membrane and $C$ the composite membrane), respectively. $\gamma$ is the Stefan-Boltzmann constant, $T_0$ is the temperature at the bottom of the silicon frame and the ambient temperature.

The general solution of Eq. (2) can be described as:

$$\widetilde{\Delta T}(x,t,2\omega) = \widetilde{\Delta T}(x,2\omega)e^{i(2\omega t+\pi)} \qquad (3)$$

where $\widetilde{\Delta T}(x,2\omega)$ is complex temperature oscillation. By inserting Eq. (3) into Eq.(2), one can get:

$$\frac{\partial^2 \widetilde{\Delta T}(x,2\omega)}{\partial x^2} - \mu(2\omega)^2 \widetilde{\Delta T}(x,2\omega) = 0 \qquad (4),$$

where $\mu(2\omega)^2$ is given by:

$$\mu(2\omega)^2 = i\frac{2\omega}{\alpha_{\|,j}} + \frac{8\varepsilon_j \gamma T_0^3}{\kappa_{\|,j}d_j} \qquad (5).$$

The boundary conditions of Eq. (4) are given by:

$$\widetilde{\Delta T}(x,2\omega)|_{x=l/2} = 0 \qquad (6)$$

and

$$-\kappa_j d_j L_h \frac{d\widetilde{\Delta T}(x,2\omega)}{dx}\bigg|_{x=0} = \frac{P_0}{2} - i2\omega C_T \widetilde{\Delta T}(x,2\omega)\big|_{x=0} \quad (7)$$

where $l/2$ is the half-width of the suspended membrane, $L_h$ is the length of S1; $C_T$ represents the heat capacity of S1 and the membrane structure beneath it (hatched area in Fig. 1b) [15]:

$$C_T = C_{V,h} L_h d_h \frac{w_h}{2} + C_{V,j} L_h d_j \frac{w_h}{2} \quad (8)$$

where $C_{V,h}$、$d_h$、$w_h$ are volumetric heat capacity、thickness and width of the metal strip，$C_{V,j}$ is the volumetric heat capacity of the membrane structure $j$ （$j = M, C$）.

The final solution of Eq. （4） is found to be：

$$\widetilde{\Delta T}(x,2\omega) = \frac{(P_0/2)\cdot\sinh[\mu(2\omega)(l/2-x)]}{\kappa_{\parallel,j} d_j L_h \mu(2\omega)\cdot\cosh[\mu(2\omega)l/2] + i2\omega C_T \cdot\sinh[\mu(2\omega)l/2]} \quad (9).$$

Equation (9) has a similar form to the general solution of the 1D heat diffusion equation given by Sikora and co-workers. [15] The imaginary term （heat capacity term） on the denominator of Eq. (9) can be dropped off if it is much smaller than the real term:

$$\widetilde{\Delta T}(x,2\omega) = \frac{P_L \cdot \sinh[\mu(2\omega)(l/2-x)]}{2\kappa_{\parallel,j} d_j \mu(2\omega)\cdot\cosh[\mu(2\omega)l/2]} \quad (10).$$

where $P_L$ is the heating power per length dissipated from S1. Equations (9) and (10) are hereafter referred to as 1D "full" model and finite width (FW) model, respectively.

To facilitate the data reduction, one may assume that the suspended membrane is semi-infinitely wide, which is valid if the heating frequency $2\omega$ is high enough to satisfy the condition of $l/2 \gg \mu(2\omega)^{-1}$. Thus Eq.

(10) can be further simplified as：

$$\widetilde{\Delta T}(x,2\omega) \approx \frac{P_L}{2\kappa_{\parallel,j}d_j\mu(2\omega)}e^{-\mu(2\omega)x} \quad (11)$$

For the suspended membrane structure, the radiation heat loss at room temperature is often considered to be negligible [12, 16]. Thus, neglecting the term of the radiation heat loss in Eq. (5) leads to:

$$\mu(2\omega)^2 = i\frac{2\omega}{\alpha_{\parallel,j}} \quad (12).$$

By substituting Eq. (12) into Eq. (11), we can get spatially varied amplitude and phase shift of the temperature oscillation, which are defined as MPS model in this paper:

$$|\Delta T(x,2\omega)|_{MPS} = \frac{P_L}{2\sqrt{2}\kappa_{\parallel,j}d_j}\sqrt{\frac{\alpha_{\parallel,j}}{\omega}}\exp\left(-\sqrt{\frac{\omega}{\alpha_{\parallel,j}}}x\right) \quad (13),$$

$$\Phi(x,2\omega)_{MPS} = -\sqrt{\frac{\omega}{\alpha_{\parallel,j}}}x + \frac{5}{4}\pi \quad (14).$$

For the MPS method, the amplitude on S1 ($|T(2\omega)|_{S1,MPS}$) and the phase shift on S2 ($\Phi(2\omega)_{S2,MPS}$) are used to extract $\alpha_{\parallel,j}$ and $\kappa_{\parallel,j}$ ($j=M,C$) [12, 13].

Recently, Zhu proposed a modified Ångström method for measuring the thermal conductivity of strip-like bulk samples, which takes the radiation and convection heat loss into consideration [14]. Since the method is also based on the periodic solution of the 1D heat diffusion equation, it is possible to apply it to the suspended membrane configuration discussed here. Unlike the conventional Ångström method, whose experimental setup consists of one independent heater and two thermometers (*e.g.*, thermocouples or infrared thermometers), the membrane-based modified Ångström (MMA) method proposed here is compatible with the DMS configuration by merging the heater with one of the thermometer.

Following Zhu's treatment, we can rewrite Eq. (5) as：

$$\mu(2\omega) = M(2\omega) + iN(2\omega) = \sqrt{M(2\omega)^2 + N(2\omega)^2}\, e^{i\varphi_0} \quad (15),$$

where

$$\varphi_0 = \tan^{-1}\left(N(2\omega)/M(2\omega)\right) \quad (16).$$

Note that Eq. (15) contains the radiation heat loss term.

By introducing Eq. (15), Eq. (11) is transformed into the following form (denoted as MMA model)：

$$\widetilde{\Delta T}(x, 2\omega) = \frac{P_L}{2\kappa_{\|,j} d_j} \frac{e^{-M(2\omega)x}}{\sqrt{M(2\omega)^2 + N(2\omega)^2}} \sin\left(2\omega t + \Phi(x, 2\omega)_{MMA}\right) \quad (17)$$

where $\Phi(x, 2\omega)_{MMA}$ is spatially varied phase shift of the temperature oscillation defined in the MMA model:

$$\Phi(x, 2\omega)_{MMA} = -N(2\omega)x + \frac{3}{2}\pi - \varphi_0 \quad (18).$$

From Eq. (16) and Eq. (18), it is straightforward to show that:

$$M(2\omega) = N(2\omega)/\tan\left(\frac{3}{2}\pi - \Phi(2\omega)_{S1,MMA}\right) \quad (19)$$

and

$$N(2\omega) = \frac{\Phi(2\omega)_{S1,MMA} - \Phi(2\omega)_{S2,MMA}}{D} \quad (20),$$

where $\Phi(2\omega)_{S1,MMA}$ and $\Phi(2\omega)_{S2,MMA}$ are the phase shift on S1 and S2, respectively.

On the other hand, the amplitude of the temperature oscillation on S1 defined in the MMA model, $|\Delta T(2\omega)|_{MMA,S1}$, is given by：`

$$|\Delta T(2\omega)|_{S1,MMA} = \frac{P_L}{2\kappa_{\|,j} d_j} \frac{1}{\sqrt{M(2\omega)^2 + N(2\omega)^2}}, \quad (21)$$

For the MMA method, one can get a series of $\left(\sqrt{M(2\omega)^2 + N(2\omega)^2}\right)^{-1}$ and $|\Delta T(2\omega)|_{S1}$ by varying ω. From the slope of the $\left(\sqrt{M(2\omega)^2 + N(2\omega)^2}\right)^{-1}$ vs.

$\left|\Delta T(2\omega)\right|_{S1}$ plot, ie. $2\kappa_{\parallel,j}d_j/P_L$, the in-plane thermal conductivity $\kappa_{\parallel,j}$ ($j=M,C$) can be directly extracted [14].

The in-plane thermal conductivity of the test film $\kappa_{\parallel,f}$ is calculated from:

$$\kappa_{\parallel,f} = \left(\kappa_{\parallel,C}d_C - \kappa_{\parallel,M}d_M\right)/d_f \qquad (22)$$

In order to reliably characterize the thin film with low $\kappa_{\parallel,f}$ (e.g., < 1 W m$^{-1}$ K$^{-1}$), $d_f$ should be large enough to ensure that the thermal conductance of the test film is comparable or larger than that of the dielectric membrane [13], and $\kappa_{\parallel,M}$ of the dielectric membrane needs to be determined as accurately as possible.

## 2.2 FEM simulations

Three-dimensional (3D) FEM simulations have been carried out by employing the COMSOL™ Multiphysics software to obtain the temperature distributions on the suspended membrane structures under periodic electrical heating in vacuum. The simulations were achieved by coupling the heat transfer module with the electric current module embedded in the COMSOL™ software. The heat radiation from the chip surface is modeled at ambient temperature (T$_{amb}$) by applying the Stefan-Boltzmann law of blackbody radiation. The material properties of the silicon frame and the gold electrodes used for the FEM modellings and 1D calculations are listed in Table 1, while those of the SiN$_x$ membrane and four test films are listed in Table 2. All input parameters mentioned above are assumed to be temperature independent. Note that the film_A represents a typical conductive polymer material whose thermophysical properties are chosen from the values of a PEDOT:PSS sample reported by Wei and co-workers[17].

|  | Gold | Silicon |
|---|---|---|
| $\kappa$ (W m$^{-1}$ K$^{-1}$) | 240.00 * | 142.20 |
| $C_V$ (J m$^{-3}$ K$^{-1}$) | $1.60 \times 10^6$ * | $1.63 \times 10^6$ |
| $\varepsilon$ | 0.03 ** | 0.5 |
| $\sigma$ (S m$^{-1}$) | $4.7 \times 10^7$ | / |

*Ref[11];   **Ref[10]

Table 1.   Materials properties of gold and silicon used in the simulations

|  | SiN$_x$ | Film_A | Film_B | Film_C | Film_D |
|---|---|---|---|---|---|
| $\kappa_\parallel$ (W m$^{-1}$ K$^{-1}$) | 3.00 | 0.84 * | 10.00 | 30.00 | 50.00 |
| $\alpha_\parallel$ (m$^2$ s$^{-1}$) | $1.15 \times 10^{-6}$ | $6.00 \times 10^{-7}$ * | $7.14 \times 10^{-6}$ | $2.14 \times 10^{-5}$ | $3.57 \times 10^{-5}$ |
| $C_V$ (J m$^{-3}$ K$^{-1}$) | $2.60 \times 10^6$ | $1.40 \times 10^6$* | $1.40 \times 10^6$ | $1.40 \times 10^6$ | $1.40 \times 10^6$ |
| $\varepsilon$ | 0.3 | 0.6 | 0.6 | 0.6 | 0.6 |
| $d$ (nm) | 100 | 300 | 1000 | 1000 | 1000 |

*Ref[17]

Table 2.   Material properties and thickness of SiN$_x$ membrane and test films used in the simulations.

The amplitude of injecting current ($I_0$) is adjusted in such a way that the amplitude of the temperature oscillation on S1 is smaller than 5.0 K in the low frequency side (*e.g.*, 10 Hz), and the amplitude of the temperature oscillation on S2 at 250 Hz can reach at least several tens of millikelvin (see section 3.1). In fact, $I_0$ needs to be raised accordingly with the increase of $\kappa_{\parallel,C}$ which represents essentially an effective in-plane thermal conductivity of the double-layered membrane. In Table 3, $\kappa_{\parallel,C}$ and $I_0$ are listed together with $\alpha_{\parallel,C}$ for each composite film.

|  | SiN$_x$/Film_A | SiN$_x$/Film_B | SiN$_x$/Film_C | SiN$_x$/Film_D |
| --- | --- | --- | --- | --- |
| $\kappa_{\parallel,C}$ (W m$^{-1}$ K$^{-1}$) | 1.38 | 9.36 | 27.55 | 45.72 |
| $\alpha_{\parallel,C}$ (m$^2$ s$^{-1}$) | 8.12 × 10$^{-7}$ | 6.20 × 10$^{-6}$ | 1.83 × 10$^{-5}$ | 3.03 × 10$^{-5}$ |
| $I_0$ (mA) | 1.0 | 1.3 | 1.6 | 2.0 |

Table 3. In-plane thermal conductivity 、in-plane thermal diffusivity and the amplitude of the applied input electrical current for the composite membranes.

Figure 2 shows an instantaneous temperature distribution plot at t = 0.16s for the bare SiN$_x$ membrane heated by an electrical current of $I_0$ = 0.7 mA and $f$ =100 Hz ( $f = \omega/2\pi$ ) at an ambient temperature of 300K.

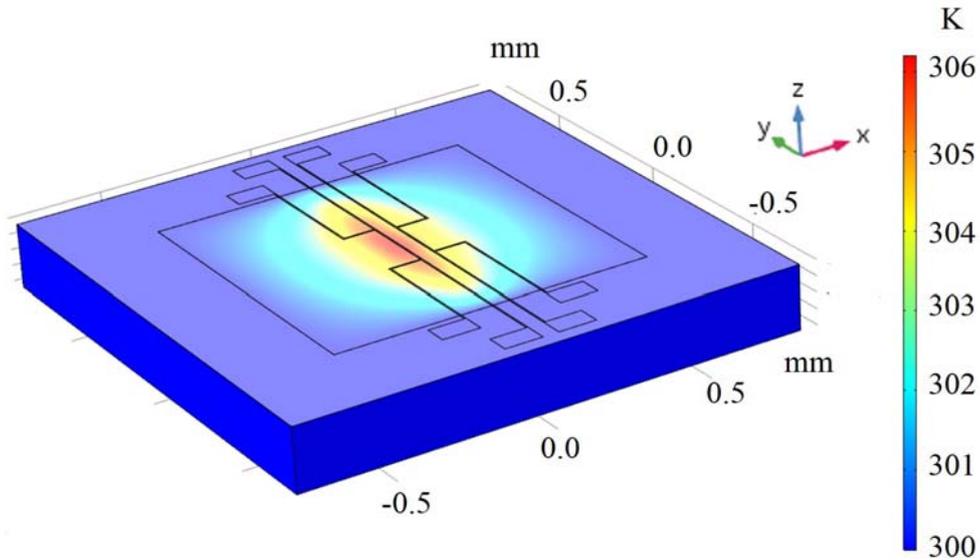

Fig. 2 Instantaneous temperature distribution （t = 0.16 s）calculated by the FEM simulation for the bare SiN$_x$ membrane heated by an electric current of 0.7 mA and a frequency of 100 Hz at T$_{amb}$=300K.

The spatially averaged temperature oscillations on S1 and S2 are plotted as a function of time in Fig. 3. The amplitude and phase shift of the temperature oscillations can be extracted by fitting the steady portion of the curves to a sinusoidal function. The values are used to mimic experimentally measured observables, from which we can reversely solve the in-plane thermal conductivity of the membrane structure through the MPS and MMA methods. The results are employed to compare with the corresponding preset values of the FEM simulations to evaluate the reliability and application scope of the two methods.

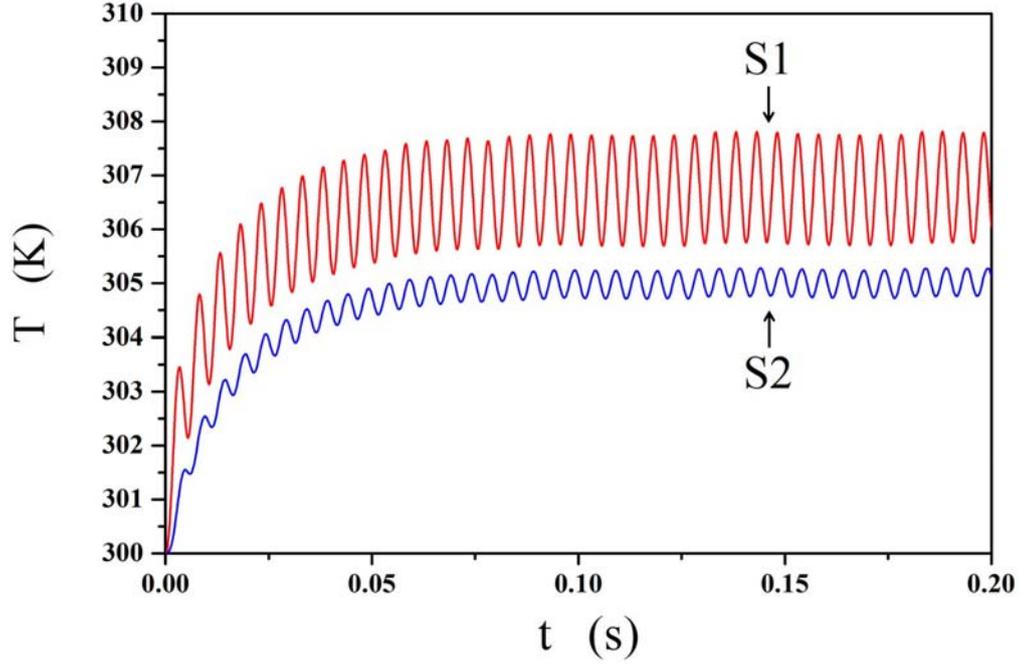

Fig. 3 Spatially averaged temperature oscillations on S1 (red) and S2 (blue) as a function of time calculated by the FEM simulation for the bare SiN$_x$ membrane ($I_0$ = 0.7 mA, f = 100 Hz, $T_{amb}$=300K).

## 3. Results and Discussions

## 3.1 Extracting $\kappa_{\parallel,M}$ of the bare SiN$_x$ membrane using the MPS method

To imitate the experimental process of extracting $\kappa_{\parallel,j}$ ($j=M,C$) from the FEM results using the MPS or MMA method, it is necessary to determine the frequency range of the heating current used for the data fitting. As discussed in section 2, both the MPS method and MMA method are based on the simplified 1D models. In principle, the trend of the frequency dependent temperature rises on S1 and S2 predicted by these 1D models should be as close as possible to that occurring on the 3D membrane structure. To fulfill this requirement, the following conditions should be satisfied: (1) There is negligible heat flow in the out-of-plane direction of the membrane structure [18]; (2) The heating strip S1 is infinitely narrow [18]; (3) S1 is infinitely long and there is negligible side heat flow starting from S1 towards the chip frame along the metal bridges and the membrane structures [16]; (4) The membrane structure is semi-infinitely wide. To meet the conditions (1) and (2), the cross-plane thermal penetration depth $L_{D,\perp}$ and the in-plane thermal penetration depth $L_{D,\parallel}$ (The thermal penetration depth characterizes the distance that heat diffuses in the materials during one cycle of heating period) should satisfy:

$$L_{D,\perp} = \sqrt{\frac{\alpha_{\perp,j}}{\omega}} = 5d_j \geq d_j \qquad (23)$$

$$L_{D,\parallel} = \sqrt{\frac{\alpha_{\parallel,j}}{\omega}} = 5\left(\frac{w_h}{2}\right) \geq \frac{w_h}{2} \qquad (24)$$

where $\alpha_{\perp,j}$ and $\alpha_{\parallel,j}$ ($j=M,C$) are the cross-plane thermal diffusivity and the in-plane thermal diffusivity of the membrane structure, respectively. $w_h/2$ and $d_j$ are the half-width of the heating strip and the thickness of membrane structure, respectively, as defined previously. The upper limit of the current frequency $f_{\max}$ is thus given by either of the following equations, depending on which one yields a smaller value:

$$f_1 = \alpha_{\perp,j} / 50\pi d_j^2 \qquad (25)$$

$$f_2 = \alpha_{\parallel,j} / 12.5\pi w_h^2 \qquad (26).$$

It can be seen from Fig. 1 that S1 and S2 are connected through metal bridges to the connection pads located on a $SiN_x$ region right above the silicon frame which acts as a heat sink. When S1 is heated, there will be a temperature drop along the metal bridges. To meet the condition of the infinitely long heating strip, the thermal conductance from one of the sides of the heating stripe S1 along +x or –x direction, $G_j$ ($j=M,C$), should be much greater than the thermal conductance along the metal bridges $G_b$ [16]:

$$G_j = \frac{\kappa_{\parallel,j} L_h d_j}{L_{D,\parallel}} \gg G_b = \kappa_h \left( \frac{w_b d_b}{L_b} + \frac{w_v d_b}{L_v} \right) \quad (27)$$

where $\kappa_h$、$w_b$、$w_v$、$L_b$、$L_v$ and $d_b$ are the thermal conductivity of the metal strip S1, the width of the current-injection bridge, the width of the voltage-measurement bridge, the distance between S1 end and the contact pad along the current-injection bridge, the distance between S1 end and the contact pad along the voltage-measurement bridge and the thickness of the bridges, respectively. If we assume that $G_j = 5G_b$, a frequency limit can be found to be:

$$f_3 = \frac{\alpha_{\parallel,j}}{2\pi} \left[ \frac{5\kappa_h}{\kappa_{\parallel,j} L_h d_j} \left( \frac{w_b d_b}{L_b} + \frac{w_v d_b}{L_v} \right) \right]^2 \quad (28)$$

Finally, to fulfill the condition (4), the in-plane thermal penetration depth $L_{D,\parallel}$ should be smaller than the half width of the suspended membrane $l/2$,

$$L_{D,\parallel} = \sqrt{\frac{\alpha_{\parallel,j}}{\omega}} = \frac{1}{5}\left(\frac{l}{2}\right) \leq \frac{l}{2} \quad (29)$$

which leads to:

$$f_4 = 50\alpha_{\parallel,j} / \pi l^2 \quad (30),$$

The lower limit of the current frequency is determined by either $f_3$ or $f_4$, depending on which one is larger. One can use typical values of the thermal conductivity and the thermal diffusivity of dielectric layers to estimate the frequency range for determining $\kappa_{\parallel,M}$ of the bare $SiN_x$ membrane based on the chip configuration shown in Fig. 1a. By plugging $\alpha_{\parallel,M} = \alpha_{\perp,M} = 10^{-6}$ m$^2$s$^{-1}$, $\kappa_{\parallel,M} = 2$ W m$^{-1}$ K$^{-1}$, $\kappa_h = 250$ W m$^{-1}$ K$^{-1}$, $L_h = 0.25$ mm, $d_M = 100$ nm, $d_b = 40$ nm, $w_b = w_v = w_h = 5$ μm, $L_b = 0.37$ mm, $L_v = 0.47$ mm, $l = 1$ mm into Eq. (25), Eq. (26), Eq. (28) and Eq. (30), $f_1$, $f_2$, $f_3$ and $f_4$ Hz are found to be 6.4 × 10$^5$ Hz、1.0 × 10$^3$ Hz、86 Hz and 8.0 Hz，respectively. Thus the frequency range for characterizing the 100 nm thick $SiN_x$ membrane is roughly between 80 Hz and 1000 Hz, which is dominated by the width of heating strip and the side heat flow through the metal bridges.

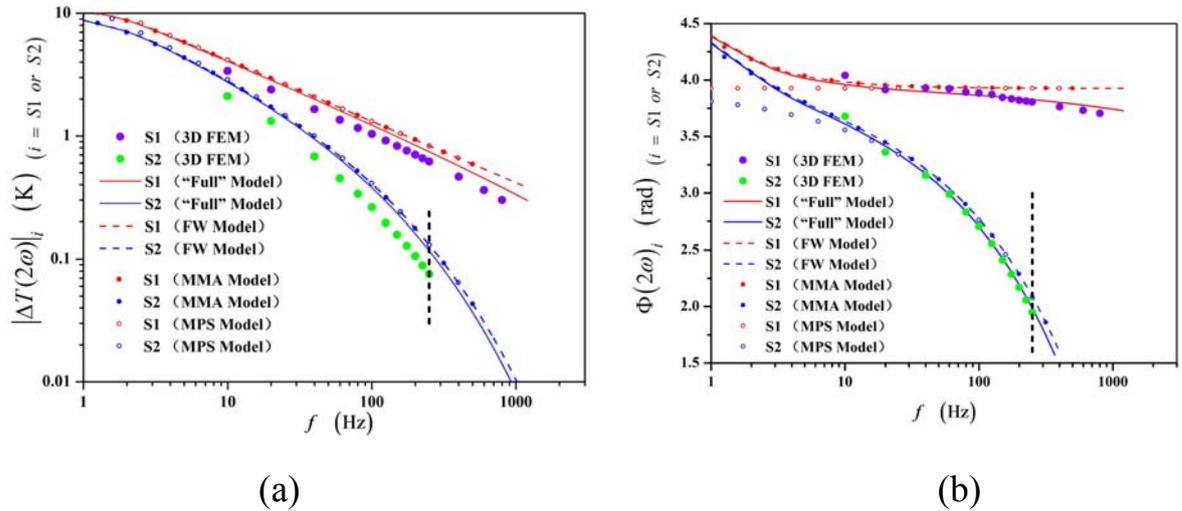

(a)  (b)

Fig. 4  Amplitude (a) and phase shift (b) of the temperature oscillation on S1 and S2 for a 100 nm thick $SiN_x$ membrane as a function of the heating current frequency, calculated by the FEM modeling and the 1D analytical solutions. The vertical lines mark the upper frequency limit of extracting the FEM results from the S2 strip.

The frequency-dependent amplitude and phase shift of the temperature oscillation on S1 and S2 for the SiN$_x$ membrane have been calculated at T$_{amb}$ = 300 K by the FEM modeling. The results are displayed in Fig. 4 in comparison with predictions of the 1D analytical models. As shown in Fig. 4(a), the amplitude of the temperature oscillations on S1 and S2 (*i.e.*, $|\Delta T(2\omega)|_{S1}$ and $|\Delta T(2\omega)|_{S2}$) undergo rapid decline with frequency, and $|\Delta T(2\omega)|_{S2}$ decays even faster in the high frequency range. Since the low $|\Delta T(2\omega)|_{S2}$ value may cause measurement difficulties experimentally, an upper frequency limit of 250 Hz has been imposed，above which the temperature rise on S2 computed by the FEM simulation has been discarded. At this frequency limit, $|\Delta T(2\omega)|_{S2}$ should maintain a measureable level of at least several tens of millikelvin by tuning the magnitude of I$_0$.

Among the 1D analytical models, the 1D "full" model has considered the radiation heat loss from a finite wide membrane. It also includes a heat capacity term for S1 and the corresponding membrane structure beneath S1, which is neglected in the FW model. Both the MPS model and MMA model are simplified from the FW model by assuming that the membrane structure is semi-infinitely wide. The difference between the MMA model and the MPS model is that the latter one ignores the radiation heat loss.

Figure 4 shows that the frequency-dependent trends of $|\Delta T(2\omega)|_{S1}$、

$\left|\Delta T(2\omega)\right|_{S2}$, $\Phi(2\omega)_{S1}$ and $\Phi(2\omega)_{S2}$ derived from the 3D FEM simulations agree in general with those predicted by the four 1D models. A close examination reveals that the amplitudes and phase shifts calculated by the FW model, the MPS model and the MMA model are almost completely overlapped above 20 Hz, which implies that the radiation heat loss doses not play a significant role in this frequency region at room temperature. However, since the three models ignore the heat capacity term in Eq. (9), they yield larger $\left|\Delta T(2\omega)\right|_{S1}$ and $\left|\Delta T(2\omega)\right|_{S2}$ than the 1D "full" model and the FEM simulation, and the deviations increase continuously with frequency since the heat capacity term is proportional to frequency. (Fig. 4a) In fact, the 1D "full" model also overestimates $\left|\Delta T(2\omega)\right|_{S1}$ and $\left|\Delta T(2\omega)\right|_{S2}$ computed by the FEM simulation due to the fact that the 1D model neglects the side heat flow through the bridges.

The phase shifts above 20 Hz can be clearly divided into two groups: those calculated by the FW model、the MPS model and the MMA model, and those by the 1D "full" Model and the FEM simulation. (Fig. 4b) The $\Phi(2\omega)_{S1}$ curves in the former group stay almost constant (approximately $5\pi/4$) in the frequency range of 20–1000Hz, while those in the latter group declines slowly from ~ $5\pi/4$ between 20 to100 Hz and decays remarkably above 100 Hz. The difference between the two groups should be also attributed to the influence of the heat capacity term mentioned above. In summary, the heat capacity and the side heat flow associated with the strips and the bridges is prominent when measuring the bare SiNx membrane. In order to faithfully retrieve $\kappa_{\parallel,M}$ experimentally, the strips and the bridges should be made as narrow and thin as possible. It is worth mentioning that the phase shift below 10 Hz predicted by the MPS model differs significantly from that forecasted by the other three 1D models, which implies that the heat radiation from the surface of the thin SiNx membrane is remarkable in the low frequency limit.

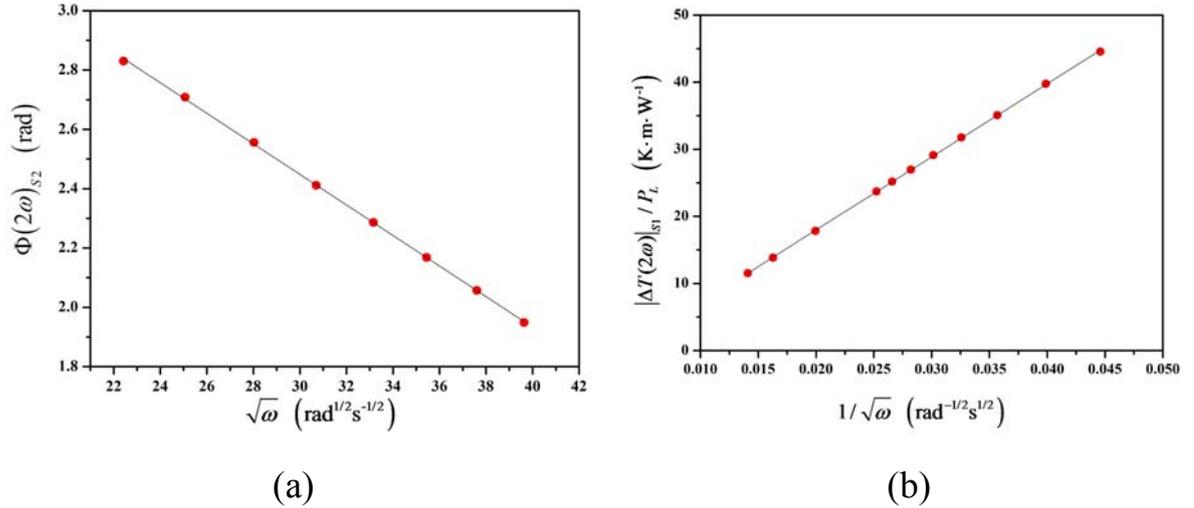

(a)                                    (b)

Fig.5 Determination of the in-plane thermal conductivity of SiNx membrane from the FEM data display in Fig.4 using the MPS method: (a) plot of $\Phi(2\omega)_{S2}$ as a function of $\sqrt{\omega}$; (b) plot of $|\Delta T(2\omega)|_{S1}/P_L$ as a function of $1/\sqrt{\omega}$. The solid line in (a) and (b) corresponds to the linear least-squares fitting to Eq. (14) and Eq. (13), respectively.

Figure 5 illustrates the procedure of employing the MPS method to extract $\kappa_{\parallel,M}$ of the bare SiNx membrane from the FEM data shown in Fig. 4. The in-plane thermal diffusivity is determined by linearly fitting $\Phi(2\omega)_{S2}$ versus $\sqrt{\omega}$ to Eq. (14) in the frequency range of 150 - 250 Hz (Fig. 5a). As a result, $\alpha_{\parallel,M}$ was found to be $9.34 \times 10^{-7}$ m$^2$ s$^{-1}$, which is about 18.8% smaller than the corresponding preset value shown in Table 2. The in-plane thermal conductivity $\kappa_{\parallel,M}$ was subsequently determined to be 3.11 W m$^{-1}$ K$^{-1}$ from the slope of the linear $|T(2\omega)|_{S1} \sim 1/\sqrt{\omega}$ curve (see Eq.(13)) in the range of 150- 800 Hz, which is 3.7% larger than the true value (Fig. 5b).

### 3.2 Determining $\kappa_{\parallel,f}$ of test films using the MPS method

Figure 6 displays the amplitudes and the phases of the temperature oscillation on S1 and S2 calculated by the FEM simulations for the three composite membranes, namely, SiN$_x$/Film_A, SiN$_x$/Film_C and SiN$_x$/Film_D. The corresponding curves predicted by the 1D models are also plotted together for comparison. As shown in Fig. 6a, the $|\Delta T(2\omega)|_{S1}$ and $|\Delta T(2\omega)|_{S2}$ results for the SiN$_x$/Film_A membrane (with a preset $\alpha_{\parallel,C}$ value of $8.12 \times 10^{-7}$ m$^2$ s$^{-1}$) agree in general with those calculated by the four 1D models above 5 Hz. As $\alpha_{\parallel,C}$ increases, the lower limit of the frequency range in which the FEM modeling agrees with the 1D models shifts upwards. For the SiN$_x$/Film_C membrane (with a preset $\alpha_{\parallel,C}$ value of $1.83 \times 10^{-5}$ m$^2$ s$^{-1}$) and SiN$_x$/Film_D membrane (with a preset $\alpha_{\parallel,C}$ value of $3.03 \times 10^{-5}$ m$^2$ s$^{-1}$), the FEM results start to deviate from the anticipations of the 1D models when the frequency is below 40 Hz and 60 Hz, respectively.

On the other hand, Figure 6b suggests that the magnitude of $\alpha_{\parallel,C}$ have a stronger impact on $\Phi(2\omega)_{S1}$ and $\Phi(2\omega)_{S2}$. The FEM simulation is in agreement with the 1D models only when the frequency is above 10 Hz for the SiN$_x$/Film_A membrane, 80 Hz for SiN$_x$/Film_C and 150 Hz for SiN$_x$/Film_D. Consequently, the frequency range of carrying out data reduction with the MPS method becomes narrower as $\alpha_{\parallel,C}$ increases. Since the upper frequency limit for extracting the in-plane thermal diffusivity is 250 Hz, the data fitting should be restricted to a frequency range of 10 Hz – 250 Hz for the SiN$_x$/Film_A membrane, 80 Hz – 250 Hz for SiN$_x$/Film_C, and 150 Hz – 250 Hz for SiN$_x$/Film_D.

Figure 7 illustrates the procedure of extracting the in-plane thermal conductivity of the composite membranes from the FEM results using the MPS method. For the SiN$_x$/Film_D membrane, $\alpha_{\parallel,C}$ was found to be $1.65 \times 10^{-5}$ m$^2$ s$^{-1}$ by fitting the $\Phi(2\omega)_{S2}$ data between 150–250 Hz, and $\kappa_{\parallel,C}$ was obtained to be 33.44 W m$^{-1}$ K$^{-1}$ by fitting the $|T(2\omega)|_{S1}$ data between 150–800 Hz. The in-plane thermal conductivity $\kappa_{\parallel,f}$ of the Film_D film is thus determined to be 36.47 W m$^{-1}$ K$^{-1}$, which is 27.1% smaller than the corresponding preset value. The large error should be attributed to the facts that the assumption of the semi-infinite wide membrane breaks down and a significant amount of heat may also flow along the y-direction through the composite membrane. To apply the MPS method on the current chip configuration, the $\alpha_{\parallel,C}$ value of the composite membrane must be smaller than that of the SiN$_x$/Film_D membrane ($3.03 \times 10^{-5}$ m$^2$ s$^{-1}$, see Table 3).

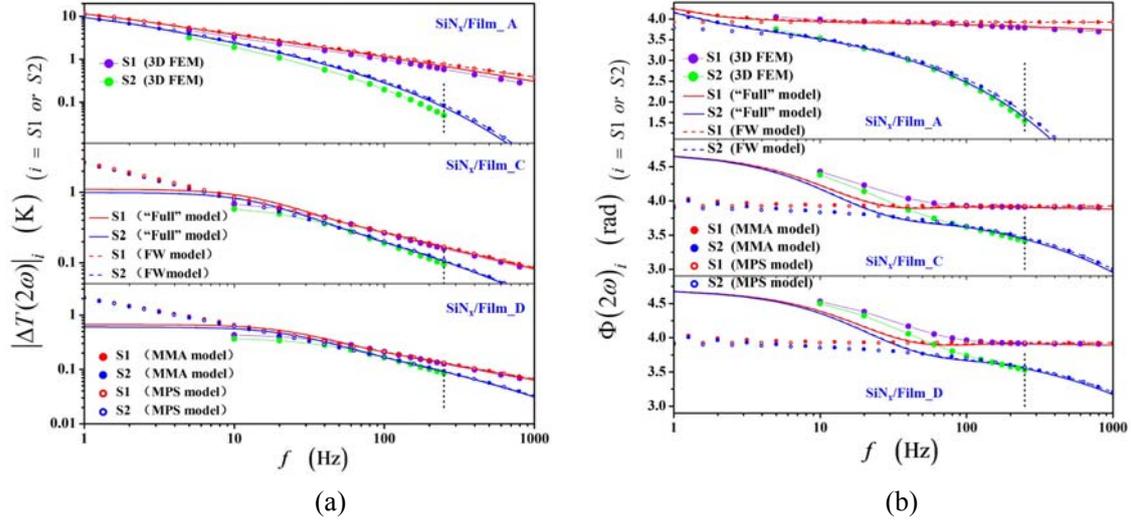

(a)　　　　　　　　　　　　　　　(b)

Fig 6. Amplitude (a) and phase shift (b) of the temperature oscillations on S1 and S2 calculated by the 3D FEM and the 1D models for the SiN$_x$/Film_A、SiN$_x$/Film_C and SiN$_x$/Film_D membranes. The vertical lines mark the upper frequency limit of extracting the FEM results from the S2 strip.

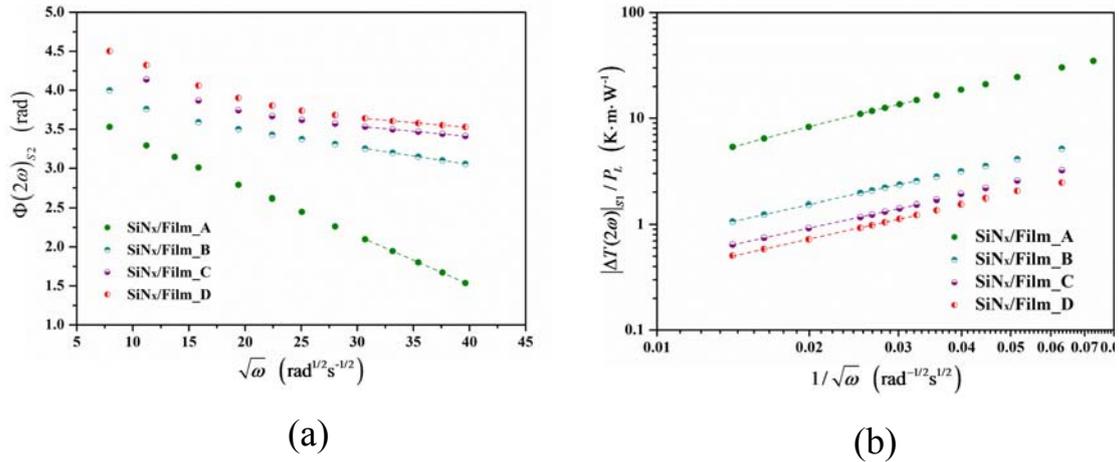

(a)　　　　　　　　　　　　　　　(b)

Fig 7. Determination of the in-plane thermal conductivity $\kappa_{\parallel,C}$ of the composite membranes from the FEM data displayed in Fig.6 using the MPS method: (a) plot of $\Phi(2\omega)_{S2}$ as a function of $\sqrt{\omega}$; (b) plot of $|T(2\omega)|_{S1}/P_L$ as a function of $1/\sqrt{\omega}$. The dashed line in (a) and (b) corresponds to the linear least-squares fitting to Eq. (14) and Eq. (13), respectively.

In actual measurement, the thermal transport properties of the test film are unknown. To be on the safe side, the frequency range of performing the data fitting should be chosen by assuming that $\alpha_{\parallel,C}$ is close to its high limit. Accordingly, the SiN$_x$/Film_A、SiN$_x$/Film_B and SiN$_x$/Film_C membranes are all fitted in the same frequency range as SiN$_x$/Film_D (Fig. 7). The derived $\kappa_{\parallel,f}$ of the four test films are summarized in Table 4. In order to evaluate the reliability of the MPS method, the relative deviation χ between the calculated $\kappa_{\parallel,f}$ and its preset value is also shown in the table. It can be seen that the χ value of the test film increases roughly with $\alpha_{\parallel,C}$ of the composite membrane. For the Film_C film, χ is found to be 8.6%. To keep the relative deviation below 8% when applying the MPS method on a chip configuration shown in Fig. 1, the upper limit of $\alpha_{\parallel,C}$ should be lower than $1.8 \times 10^{-5}$ m$^2$ s$^{-1}$. This puts a higher limit on the thickness of the test film with a high thermal conductivity. For example, to measure the gold film with the thermophysical properties listed in Table 1, it is estimated that the thickness of the film should be less than 20 nm.

|  | SiNx | Film_A | Film_B | Film_C | Film_D |
| --- | --- | --- | --- | --- | --- |
| MPS Model | 3.11 (3.7%) | 0.793 (-5.6%) | 9.64 (3.6%) | 27.42 (8.6%) | 36.47 (27.1%) |
| MMA Model | 3.09 (3.0%) | 0.842 (0.24%) | 9.80 (2.0%) | 28.77 (4.1%) | 48.10 (3.8%) |

Table 4  Comparison of the in-plane thermal conductivity (W m$^{-1}$ K$^{-1}$) derived from the MPS method and MMA method (The number in parentheses is the relative deviation χ with respect to the preset value in the FEM model)

### 3.3 Determining $\kappa_{\parallel,f}$ of test films using the MMA method

In the MPS method, two "observables", namely, $\Phi(2\omega)_{S2}$ and $|\Delta T(2\omega)|_{S1}$, are exploited to yield the in-plane thermal conductivity. By contrast, the MMA method uses three observables (i.e., $\Phi(2\omega)_{S1}$、$\Phi(2\omega)_{S2}$ and $|\Delta T(2\omega)|_{S1}$) to carry out the data reduction. Figures 8 shows the linear variation of $\left(\sqrt{M(2\omega)^2 + N(2\omega)^2}\right)^{-1}$ versus $|\Delta T(2\omega)|_{S1}$ derived from the FEM data in Fig. 4 and Fig. 6 for the bare SiN$_x$ membrane and the composite membranes within the frequency range of 80–250 Hz. Note that $\left(\sqrt{M(2\omega)^2 + N(2\omega)^2}\right)^{-1}$ is solely governed by the phase shift of the temperature oscillation, i.e., $\Phi(2\omega)_{S1}$ and $\Phi(2\omega)_{S2}$, as suggested by Eq. (19) and Eq.(20). The extracted in-plane thermal conductivity of the SiN$_x$ membrane and the test films are displayed in Table 4 to compare with the results of the MPS method. For all cases, the MMA method exhibits a smaller χ value (less than 5%) than the MPS method. The MMA method is capable to measure a composite membrane with a $\alpha_{\parallel,C}$ value of at least 3.03 × 10$^{-5}$ m$^2$ s$^{-1}$. It should be pointed out that the frequency range for data fitting is not sensitive to the magnitude of $\alpha_{\parallel,C}$. A fixed fitting frequency range of 80–250 Hz has been applied to all the membrane structures under investigation.

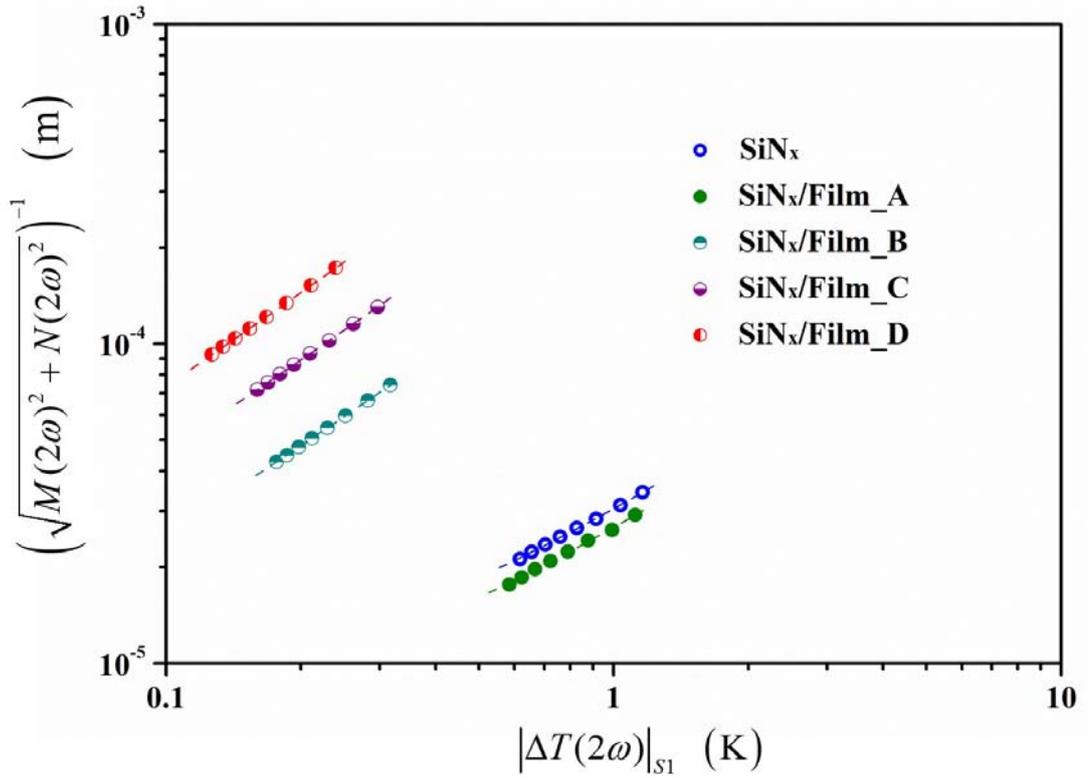

Fig. 8. Determination of the in-plane thermal conductivity of the membrane structures from the FEM data displayed in Fig.4 and Fig.6 using the MMA method: $\left(\sqrt{M(2\omega)^2+N(2\omega)^2}\right)^{-1}$ is plotted as a function of $|\Delta T(2\omega)|_{S1}$. The dashed lines are linear fits to the data.

To explain the robustness of the MMA method, we calculate the phase shift difference between S1 and S2 $\Delta\Phi(2\omega)$ ($\Delta\Phi(2\omega)=\Phi(2\omega)_{S1}-\Phi(2\omega)_{S2}$) from the FEM simulation and MMA model for different membrane structures (Fig. 9). It can be seen that the FEM simulation agrees quite well with the MMA model between 10 Hz and 250 Hz in all cases, although the idealized MMA model may fail to describe the individual $\Phi(2\omega)_{S1}$ and $\Phi(2\omega)_{S2}$ "observable" properly in the low frequency end. This suggests that the non-ideality of $\Phi(2\omega)_{S1}$ and $\Phi(2\omega)_{S2}$ associated with the FEM results may be partially cancelled out during the subtraction process. For the bare SiN$_x$ membrane, the non-ideality should mainly come from the heat capacity and the side heat flow associated with the strips and the bridges. For the composite membrane with a larger $\alpha_{\parallel,C}$, the violation of the semi-infinite width assumption is responsible for the non-ideality.

On the other hand, it can be shown from Eq. (19) and Eq. (20) that:

$$\left(\sqrt{M(2\omega)^2 + N(2\omega)^2}\right)^{-1} = \frac{D}{\Delta\Phi(2\omega)\sqrt{1 + 1/\left[\tan(3\pi/2 - \Phi(2\omega)_{S1})\right]^2}} \quad (31).$$

Equation (31) indicates that the $\left(\sqrt{M(2\omega)^2 + N(2\omega)^2}\right)^{-1}$ term is determined by $\Delta\Phi(2\omega)$ and $\Phi(2\omega)_{S1}$. It is already known from Fig. 9 that the $\Delta\Phi(2\omega)$ "observable" can be properly recovered by the MMA model. We further expect that the non-ideality associated with the "observable" pair of $|\Delta T(2\omega)|_{S1}$ and $\Phi(2\omega)_{S1}$ with respect to the MMA model, can be at least partially cancelled out when extracting the slope from the $\left(\sqrt{M(2\omega)^2 + N(2\omega)^2}\right)^{-1}$ versus $|\Delta T(2\omega)|_{S1}$ plot. Therefore, the MMA method is capable to yield relatively accurate in-plane thermal conductivity value with a rather high tolerance for selecting the frequency range of the data fitting. The MMA method, like the original modified Ångström method [14], has included the radiation heat loss term into the 1D heat diffusion equation in order to improve the measurement accuracy. Although it is actually implemented in the frequency range where the radiation heat loss is not significant, the data reduction scheme of the method does provide a self-consistent way to reduce the non-ideality of the "observables" caused by the factors such as violation of the semi-infinite width assumption.

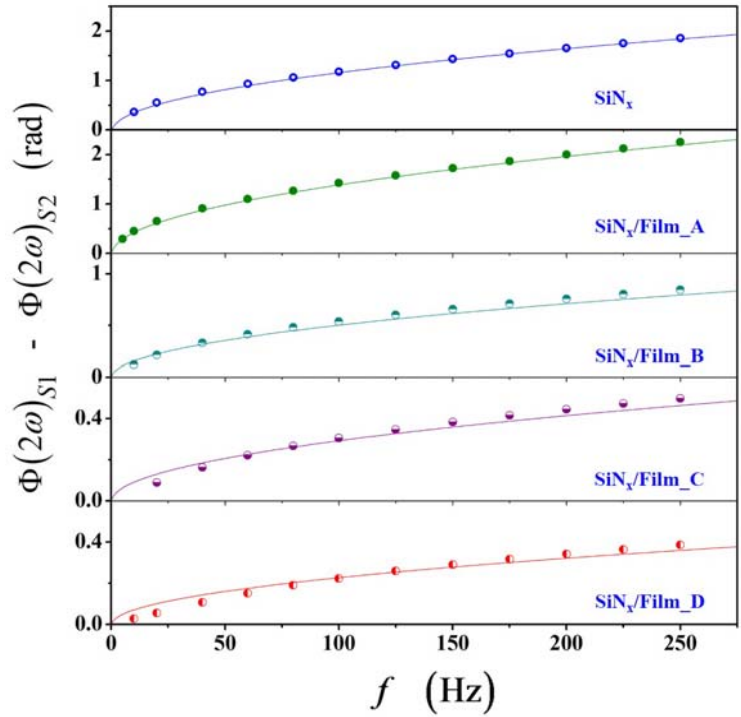

Fig. 9  $\Phi(2\omega)_{S1} - \Phi(2\omega)_{S2}$ as a function of the heating current frequency for the different membrane structures, calculated by the FEM simulations (circles) and the MMA model (solid lines).

## 4. SUMMARY

In this paper, we have conducted a series of FEM simulations and analytical calculations to evaluate the application scope of the MPS method and MMA method for measuring the in-plane thermal conductivity of thin films on a suspended membrane chip with DMS configuration (membrane size of 1 mm×1 mm and strip width of 5 μm). Compared with $|\Delta T(2\omega)|_{S1}$ and $|\Delta T(2\omega)|_{S2}$ derived from the FEM simulation, both $\Phi(2\omega)_{S1}$ and $\Phi(2\omega)_{S2}$ show more striking deviation from the MPS model and the MMA model when $\alpha_{\parallel,C}$ enhances. This is mainly due to the violation of the semi-infinite width assumption. Consequently, for the MPS method which only takes into account of the $|\Delta T(2\omega)|_{S1}$ and $\Phi(2\omega)_{S2}$ "observables", the frequency window for performing the data fitting becomes narrower as $\alpha_{\parallel,C}$ increases. For the specific membrane dimension considered here, the upper limit of $\alpha_{\parallel,C}$ should not exceed ~ 1.8 ×10$^{-5}$ m$^2$ s$^{-1}$.

On the other hand, the MMA method has made use of three "observables", namely, $|\Delta T(2\omega)|_{S1}$、$\Phi(2\omega)_{S1}$ and $\Phi(2\omega)_{S2}$. It is expected that non-ideality associated with the ($\Phi(2\omega)_{S1}$, $\Phi(2\omega)_{S2}$) pair and the ($\Phi(2\omega)_{S1}$, $|\Delta T(2\omega)|_{S1}$) pair may be at least partially cancelled out during the data reduction process. Thus the frequency range selection of performing the data fitting is not sensitive to the magnitude of the $\alpha_{\parallel,C}$ value. The upper limit of $\alpha_{C,\parallel}$ measured by the MMA method can extend to at least ~3.0 ×10$^{-5}$ m$^2$ s$^{-1}$. For typical specimen films considered in this paper whose in-plane thermal conductivity ranges from 0.84 W m$^{-1}$ K$^{-1}$ to 50 W m$^{-1}$ K$^{-1}$, the MMA method yields a theoretical measurement uncertainty of less than 5%. This study may shed some light on the development of membrane-based method of measuring the in-plane thermal conductivity of thin film.

**Acknowledgments:**

The authors would like to acknowledge the financial supports from the National Natural Science Foundation of China under grant No. 51573190 and 61176083, and the State Key Development Program for Basic Research of China under grant No. 2011CB932801.